\documentclass[english]{article}
\usepackage[T1]{fontenc}
\usepackage[latin9]{inputenc}
\usepackage{geometry}
\usepackage{color}
\usepackage{graphicx}

\makeatletter
\usepackage{times,cite}
\date{}

\@ifundefined{showcaptionsetup}{}{%
 \PassOptionsToPackage{caption=false}{subfig}}
\usepackage{subfig}
\makeatother

\usepackage{babel}
\begin{document}

\title{  Continuous variable direct secure quantum communication
using Gaussian states}

\author{  Srikara S$^{\mathparagraph,}$\thanks{Email: srikara.s@students.iiserpune.ac.in },
Kishore Thapliyal$^{\mathsection,}$\thanks{Email: kishore.thapliyal@upol.cz},
Anirban Pathak$^{\Vert,}$\thanks{Email: anirban.pathak@jiit.ac.in}
\\
{\small $^{\mathparagraph}$Indian Institute of Science Education and Research,
Pune, India}\\
{\small $^{\mathsection}$RCPTM, Joint Laboratory of Optics of Palacky University
and Institute of Physics of Academy of Science of the Czech Republic,}\\
{\small Faculty of Science, Palacky
University, 17. listopadu 12, 771 46 Olomouc, Czech Republic}\\
{\small $^{\Vert}$Jaypee Institute of Information Technology, A-10,
Sector-62, Noida, UP-201309, India}}
\maketitle
\begin{abstract}
  Continuous variable one-way and controlled-two-way secure
direct quantum communication schemes have been designed using Gaussian
states. Specifically, a scheme for continuous variable quantum secure
direct communication and another scheme for continuous variable controlled
quantum dialogue  are proposed using single-mode squeezed coherent
states. The security of the proposed schemes against a set of attacks
(e.g., Gaussian quantum cloning machine and intercept resend attacks)
has been proved. Further, it is established that the proposed schemes
do not require two-mode squeezed states which are essential for a
set of existing proposals. The controlled two-way communication scheme
is shown to be very general in nature as it can be reduced to schemes
for various relatively simpler cryptographic tasks like controlled
deterministic secure communication, quantum dialogue, quantum key
distribution. In addition, it is briefly discussed that the proposed
schemes can provide us tools to design quantum cryptographic solutions
for several socioeconomic problems.
\end{abstract}

\section{Introduction\label{sec:Introduction}}

The advantages of quantum resources and principles of quantum mechanics
in the field of quantum technology are the focus of the second quantum
revolution \cite{dowling2003quantum}. The first set of applications
of quantum mechanics was proposed in secure quantum communication
with the feasibility of a quantum key distribution protocol \cite{bennett1984quantum},
where security does not rely on the complexity of the computational
task. This was followed by several quantum key distribution and other
cryptographic schemes (see \cite{gisin2002quantum,pathak2013elements,shenoy2017quantum}
for review). Specifically, quantum resources ensure secure transmission
of a message without requiring key transmission known as quantum secure
direct communication (QSDC) \cite{bostrom2002deterministic,deng2003controlled,hu2015experimental}
and deterministic secure quantum communication (DSQC) \cite{long2007quantum}.
DSQC (QSDC) requires an (no) additional classical communication to
decode the message \cite{pathak2013elements,long2007quantum}. Bidirectional
and controlled variants of these schemes have been proposed subsequently
\cite{nguyen2004quantum,thapliyal2015applications}.

Independent of these initial discrete variable (DV) quantum cryptography
schemes, a set of continuous variable (CV) secure quantum communication
schemes have also been proposed. Information is encoded on quadratures
in CV communication schemes. The encoded information is decoded later
by homodyne or heterodyne detection (see \cite{braunstein2005quantum,andersen2010continuous,weedbrook2012gaussian}
for review). Most of the initial CV cryptography schemes used Gaussian
states, like squeezed \cite{hillery2000quantum,PhysRevA.63.022309},
Einstein-Podolsky-Rosen correlated \cite{reid2000quantum}, and coherent
\cite{ralph1999continuous,ralph2000security,grosshans2002continuous}
states. However, schemes using non-Gaussian states have also been
proposed since then (\cite{srikara2019continuous} and references
therein). CV quantum communication is preferred over corresponding
DV counterpart due to the possibility of its implementation without
requiring single photon source and/or detector, better performance
for metropolitan networks which can be performed using existing optical
communication technology.

Motivated by the advantages of CV communication in general, CV schemes
are proposed recently for QSDC \cite{pirandola2008continuous,yuan2015continuous},
DSQC \cite{marino2006deterministic}, quantum dialogue (QD) \cite{zhou2017new,zhang2019quantum},
multiparty QD \cite{pirandola2008continuous,yu2016efficient,gong2018quantum},
controlled quantum dialogue (CQD) \cite{saxena2019continuous}. Several
types of CV quantum key distribution schemes have also been designed
(\cite{srikara2019continuous} and references therein). Inspired by
these works and our recent results \cite{thapliyal2016protocols,shukla2017semi,sharma2017quantumauction,thapliyal2018orthogonal,thapliyal2019quantum,saxena2019continuous}
which established that several direct quantum communication schemes
may be useful in providing quantum solutions to secure multiparty
computation tasks \cite{yao1982protocols}, here we propose a CV QSDC
and a CV CQD schemes. Specifically, it is already established that
several socioeconomic problems can be defined as secure multiparty
computation tasks, such as voting \cite{thapliyal2016protocols},
sealed-bid auction \cite{sharma2017quantumauction}, socialist millionaire
problem \cite{saxena2019continuous}, and private comparison \cite{thapliyal2018orthogonal}.
In fact, some of the present authors have proposed a CV CQD scheme
using two-mode squeezed state \cite{saxena2019continuous} and used
it as primitive for quantum solution for the socialist millionaire
problem. {Squeezed and entangled states for quantum communication may be generated in different systems and optical processes \cite{villar2005generation,lassen2009continuous,dos2009continuous,liu2014experimental,thapliyal2014higher,thapliyal2014nonclassical,thapliyal2017comparison}.} These works and the fact that single-mode Gaussian states
are easy to prepare, justify the need for designing new schemes for
secure direct quantum communication using these states. In the rest
of the paper, we have tried to design such schemes using squeezed
coherent states.

  The rest of this paper is structured as follows. In Section
\ref{sec:Preliminaries}, we introduce some basic concepts of quantum
optics required to explain the CV QSDC and CV CQD schemes (given in
Section \ref{sec:The-Protocol}). In Section \ref{sec:Security-Analysis-of},
we analyze the security of the proposed protocols. Finally, the paper
concludes in Section \ref{sec:Conclusion}. 

\section{Preliminaries \label{sec:Preliminaries}}

  Coherent states $|\alpha\rangle$ of the radiation field
are the eigenstates of the annihilation operator $\hat{a}$ with a
complex eigenvalue $\alpha$, also called the amplitude of $|\alpha\rangle$.
A coherent state can also be described in terms of the displacement
operator $D(\alpha)$ as \cite{agarwal2013quantum}
\begin{equation}
|\alpha\rangle=D(\alpha)|0\rangle=e^{\alpha\hat{a}^{\dagger}-\alpha^{*}\hat{a}}|0\rangle.\label{eq:coh}
\end{equation}
Interestingly, the displacement operator acting on a coherent
state would give another coherent state \cite{agarwal2013quantum}
\begin{equation}
D(\beta)|\alpha\rangle=|\beta+\alpha\rangle.\label{eq:dis-coh}
\end{equation}

\begin{figure}
\begin{centering}
\subfloat[]{\includegraphics[scale=0.4]{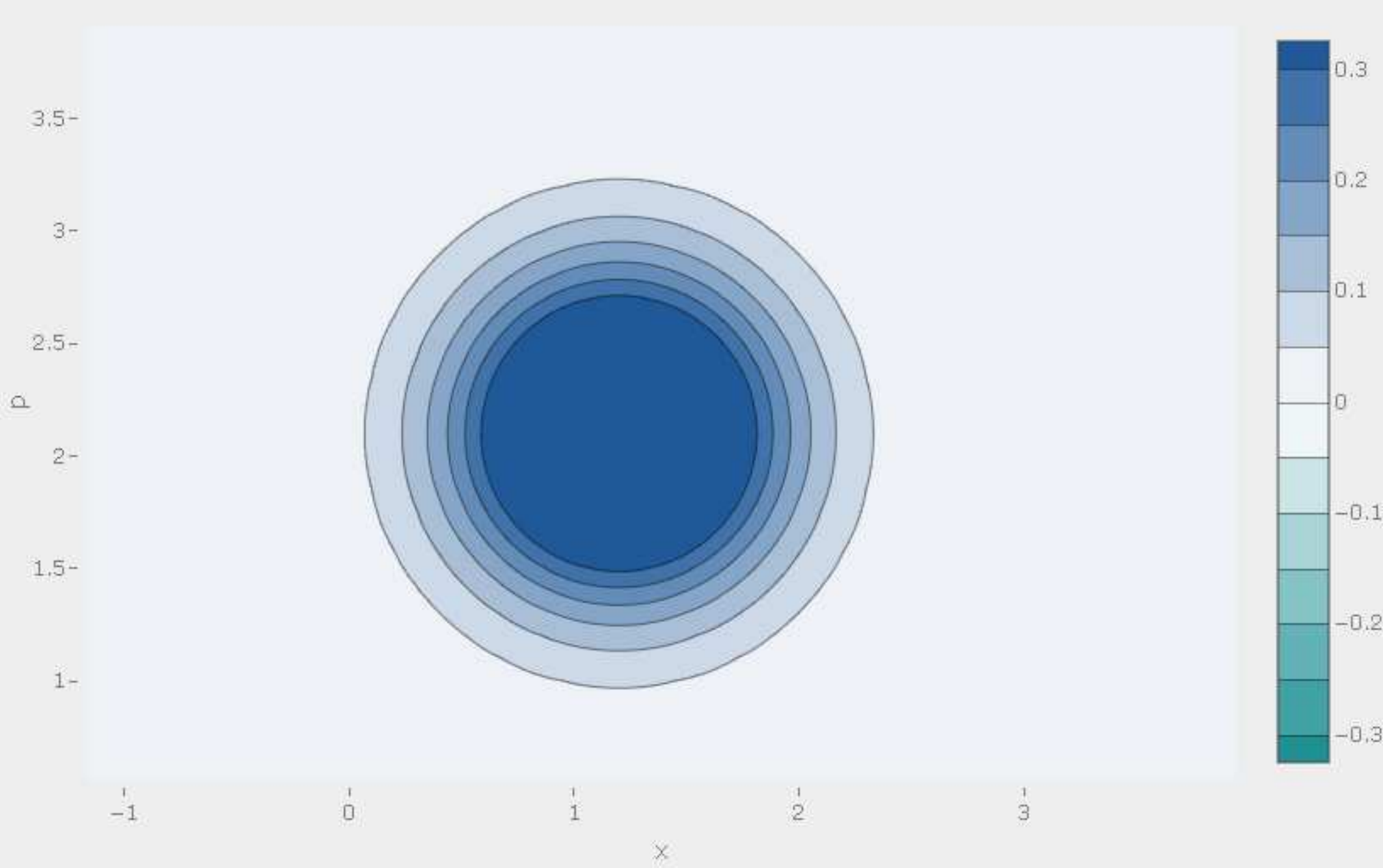}}
\subfloat[]{\includegraphics[scale=0.4]{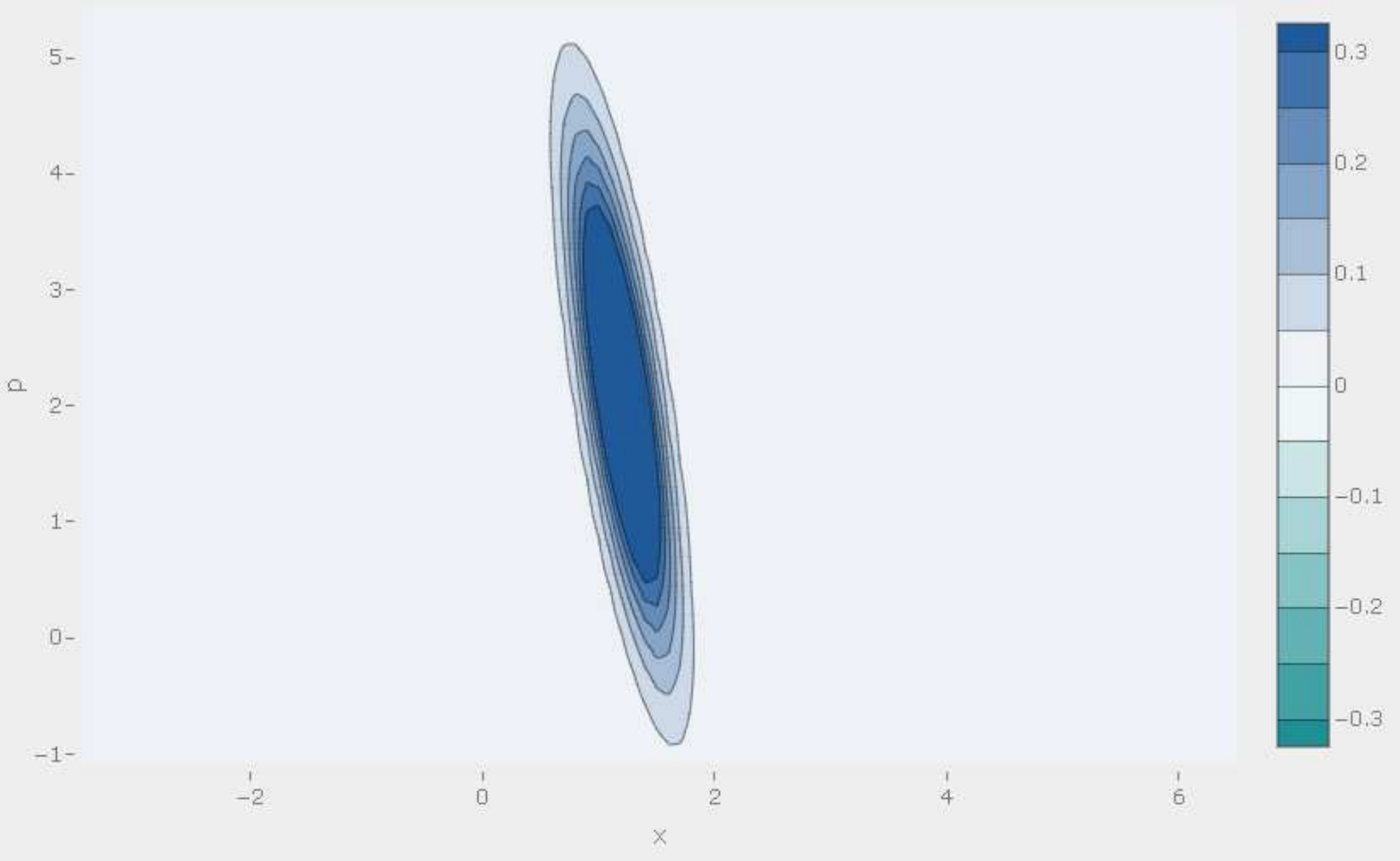}}
\par\end{centering}
\centering{}\caption{\label{Fig 1}(Color online) Wigner function of a (a) coherent state
$|\alpha\rangle$ and (b) squeezed coherent state $S(s)|\alpha\rangle$
with $s=e^{0.3i}$ with $\alpha=1.2+i2.1$. These plot are generated
using \cite{strawberryfields}.}
\end{figure}

  Further, an arbitrary quantum state $\rho$ can also be
described by a quasiprobability distribution in phase space, such
as Wigner function \cite{wigner1932quantum}
\begin{equation}
W(\gamma)=\frac{1}{\pi^{2}}\int d^{2}z\,Tr\left[\rho D(z)\right]e^{-(z\gamma^{*}-z^{*}\gamma)}.\label{eq:wig}
\end{equation}
It can be easily verified that the Wigner function of a
coherent state $|\alpha\rangle$ is obtained as a Gaussian distribution
peaked at $\gamma=\alpha$ (cf. Fig. \ref{Fig 1} (a)).

  The real part $q$ and the imaginary part $p$ of the eigenvalue
$\alpha=q+ip$ give values for the position and momentum quadratures,
respectively. These quadratures can be measured by two methods, namely
by homodyne and heterodyne detection. These methods are useful in
quantum optical tomography, too {{} \cite{thapliyal2015quasiprobability,thapliyal2016tomograms}}.
Homodyne detection is used to measure one of the quadratures, whereas
heterodyne detection helps in measuring both the quadratures. However,
Heisenberg's uncertainty principle restrict precise measurement of
both the quadratures in heterodyne measurement.

  Additionally, the squeezing operator is defined as

\begin{equation}
S(s)=e^{\frac{1}{2}(s^{*}\hat{a}^{2}-s\hat{a}^{\dagger2})},\label{eq:squ}
\end{equation}
where squeezing parameter $s=re^{i\theta}$. A coherent state gets
squeezed after application of the squeezing operator which reduces
the variance along one direction depending upon squeezing phase parameter
at the cost of increase in variance in the perpendicular direction
(cf. Fig. \ref{Fig 1} (b)). Therefore, $\alpha$ cannot be determined
accurately using homodyne/heterodyne detection in the absence of knowledge
about $s$. Note that the squeezing and displacement operators do
not commute, and thus
\begin{equation}
S(s)D(\alpha)=D(\beta)S(s),\label{eq:comm}
\end{equation}
where
\begin{equation}
\beta=\alpha\,{\rm cosh}(r)+e^{i\theta}\alpha^{*}{\rm sinh}(r)\label{eq:comm2}
\end{equation}
 and
\begin{equation}
\alpha=\beta\,{\rm cosh}(r)-e^{i\theta}\beta^{*}{\rm sinh}(r).\label{eq:comm3}
\end{equation}
 This brief description of the preliminaries enables us to propose
the protocols for CV QSDC and CV CQD that we aim to design in this
work. The next section describes the protocols.

\section{The protocols\label{sec:The-Protocol}}

  Before we describe the cryptographic schemes, we briefly
discuss the encoding scheme. The encoding scheme is as follows: The
real line is divided into 8 parts: (-$\infty$, -3), $\left(-3,-2\right),\ldots,\left(i,i+1\right),\ldots,\left(2,3\right),$
and (3, $\infty$). These parts represent the binary numbers 000,
001, 010, 011, 100, 101, 110 and 111, respectively \cite{zhou2017new,saxena2019continuous}.
The message is encoded by choosing a random real number $r$ from
the corresponding interval and applying the displacement operator
$D(\beta)$ on the incoming state where $\beta=r+ir$. Here it may
be noted that the sender need not know which quadrature is squeezed.
Consequently, she/he should encode the same message on both the quadratures. 

\subsection{Quantum secure direct communication (QSDC) protocol \label{subsec:Quantum-Secure-Direct}}

\begin{figure}
\centering{}\includegraphics[scale=1.]{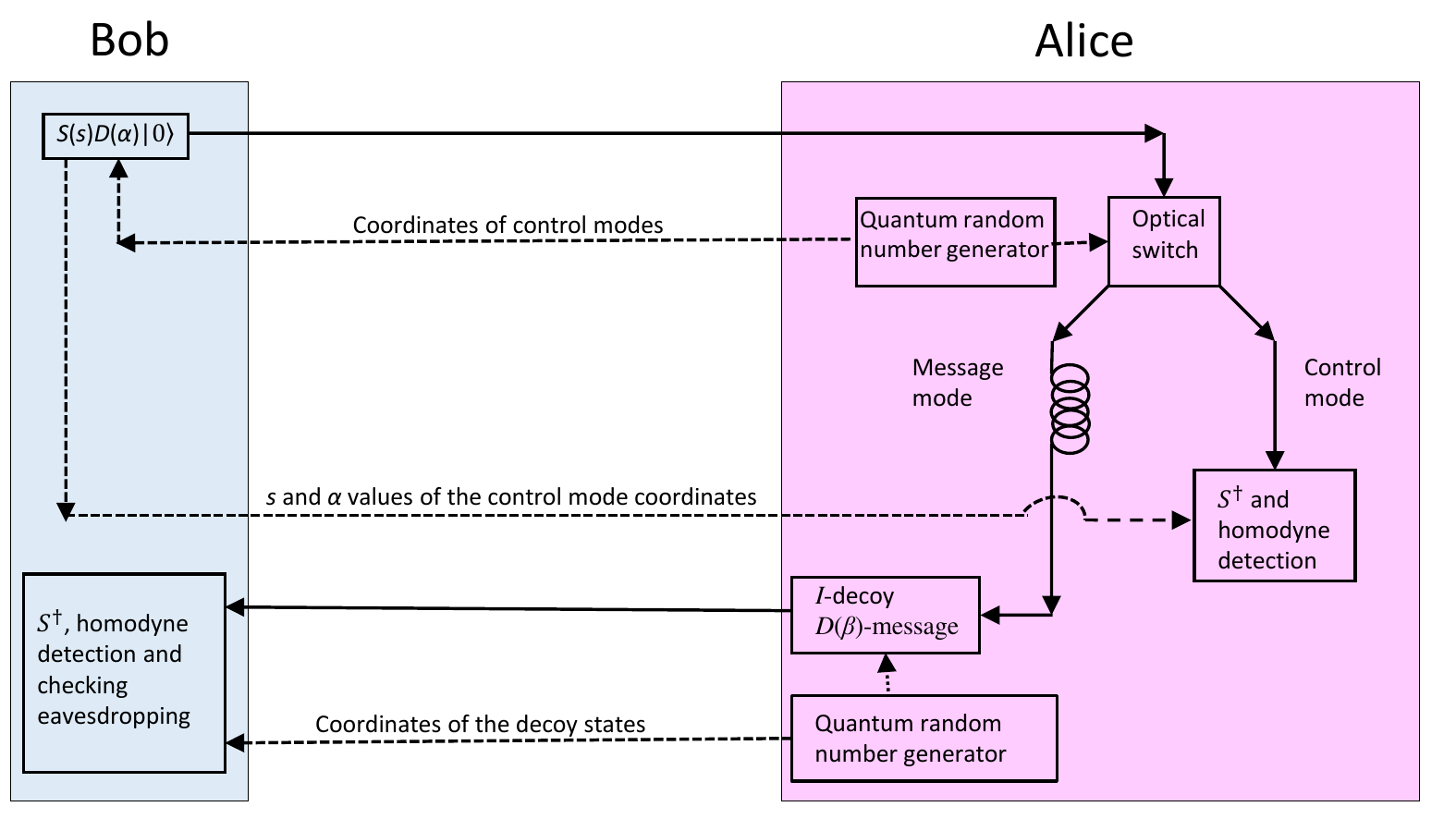}\caption{\label{Fig 3} (Color online) Schematic diagram of the QSDC protocol.}
\end{figure}

  The QSDC protocol (shown schematically in Fig. \ref{Fig 3})
works as follows
\begin{enumerate}
\item Bob generates $n$ random complex numbers in a string $R_{d}$ and
labels them $\alpha_{1},\,\alpha_{2},.....,\,\alpha_{n}$. He then
randomly chooses $n$ more complex numbers $s_{1},\,s_{2},\,....,\,s_{n}$
in a string $R_{s}$. He subsequently uses both these strings $R_{d}$
and $R_{s}$ to prepare the squeezed coherent states $S(s_{j})D(\alpha_{j})|0\rangle\,\forall j\in\left(1,\ldots n\right)$,
and sends them to Alice via block transmission.
\item Alice uses an optical switch to randomly select a set of the incoming
states as control/message mode, and sends the information about the
coordinates of the $n/2$ control mode states to Bob, whereas she
puts the message mode states into an optical delay.\\
{ Here and in what follows, by coordinates we mean the time slot chosen by Bob as control/message modes.}
\item Bob sends information of $s_{j}$ and $\alpha_{j}$ for the control
mode states to Alice. She applies $S^{\dagger}(s_{j})=S(-s_{j})$
on the corresponding control mode states, and performs homodyne measurement
in the position/momentum quadrature and verifies the real/imaginary
part of the corresponding values of $\alpha_{j}$ to check for eavesdropping.
If measured $\alpha_{j}$ values are correct up to a tolerable limit,
then she continues to Step 4, else she discards the protocol, and
they start all over again.
\item After discarding the control mode states, Alice randomly chooses $n/4$
states from the message mode and encodes her message on them by applying
$D(\beta)$, {where $\beta=m_{A}\left(1+i\right)$ depends on her message
$m_{A}\in\mathcal{R}$.} The remaining $n/4$ states are kept unchanged to be used
as decoy states, and all these $n/2$ states are then sent to Bob.
\item Bob applies the corresponding $S^{\dagger}(s_{j})$ operator and performs
homodyne measurement on the incoming $n/2$ message mode states for the
corresponding position or momentum quadrature.
\item Alice sends information about the coordinates of the decoy states
to Bob, and he checks for eavesdropping by comparing the corresponding
$\alpha_{j}$ values for the decoy states by following the same procedure
as described in Step 3.
\item The measurement results from the remaining message mode states would
reveal the message $m_{A}$ and hence Bob would obtain the message
sent by Alice using Eq. (\ref{eq:comm3}).
\end{enumerate}

\subsection{Controlled quantum dialogue (CQD) protocol \label{subsec:Controlled-Quantum-Dialogue}}

\begin{figure}
\centering{}\includegraphics[scale=1.7]{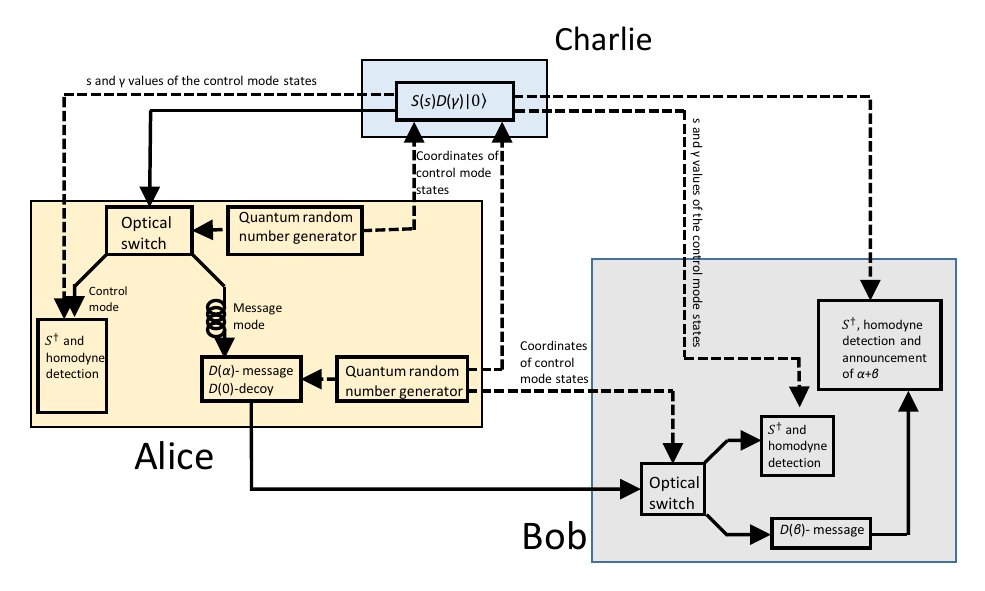}\caption{\label{Fig 3-1} (Color online) Schematic diagram of the CQD protocol.}
\end{figure}

  The CQD protocol (shown schematically in Fig. \ref{Fig 3-1})
works as follows
\begin{enumerate}
\item Charlie generates $4n$ random complex numbers as a string $R_{d}$
and labels them $\gamma_{1},\,\gamma_{2},....,\,\gamma_{4n}$. He
then randomly generates $4n$ more complex numbers $s_{1},\,s_{2},....,\,s_{4n}$
as a string $R_{s}$ and uses both $R_{d}$ and $R_{s}$ to prepare
the squeezed coherent states $S(s_{j})D(\gamma_{j})|0\rangle\,\forall j\in\left(1,\ldots4n\right)$
and sends them to Alice. 
\item Alice, using an optical switch, randomly selects $2n$ of the incoming
states as the control mode and sends the information about the coordinates {(i.e., corresponding time slots)}
of the $2n$ control mode states to Charlie, while she puts the rest
of the $2n$ states into an optical delay for the message mode.
\item Charlie sends values of $\gamma_{j}$ and $s_{j}$ of the control
mode states to Alice, and she applies the corresponding squeezing
operator $S^{\dagger}(s_{j})=S(-s_{j})$ to each of the control mode
states to perform homodyne measurement of position/momentum quadrature
on them. From the measured values she verifies the corresponding value
$\gamma_{j}$ chosen by Charlie to check for eavesdropping. If $\gamma_{j}$
values are correct up to a tolerable limit, then she continues to
the next step, else the protocol is aborted and they start all over
again.
\item After discarding the control mode states, Alice encodes her message
$m_{A}$ randomly in $n$ of her message mode states by applying the
appropriate displacement operator $D(\alpha_{j})$ with $\alpha_{j}=m_{A}+i\:m_{A}$
on them and leaves the other $n$ states as control mode. Subsequently,
she chooses a random complex number $w_{j}$ and squeezes all the
$2n$ states by applying $S(w_{j})$ to these states and sends them
to Bob. Upon Bob's confirmation of receiving all the $2n$ states,
Alice sends Bob the value of $w_{j}$ and the coordinates of the control
mode states and Charlie sends Bob their corresponding $\gamma_{j}$
and $s_{j}$.\\
Without loss of generality, we can assume $w_{j}=w\,\forall j$ for
squeezing operations would be sufficient in providing security against
participant attack by Charlie. 
\item Bob first applies the squeezing operator $S^{\dagger}(w_{j})=S(-w_{j})$
on all the $2n$ received states and then applies the corresponding
squeezing operators $S^{\dagger}(s_{j})$ on the $n$ control mode
states and checks for eavesdropping just as Step 3, meanwhile he encodes
his message $m_{B}$ in the other $n$ message modes by applying the
appropriate displacement operator $D(\beta_{j})$ with $\beta_{j}=m_{B}+i\,m_{B}$
on them. 
\item Once Bob encodes the message in the message mode states and confirms
that there is no eavesdropping, then Charlie sends the corresponding
$\gamma_{j}$ and $s_{j}$ of the message mode states, and Bob applies
the corresponding operator $D^{\dagger}(\gamma_{j})S^{\dagger}(s_{j})$
on the message mode states. He measures position/momentum quadrature
using homodyne detection and announces $\alpha_{j}+\beta_{j}$ after
calculating it using Eq. (\ref{eq:comm3}). \\
Finally, Alice and Bob can obtain each other's message by subtracting
their message from the announced value of $\alpha_{j}+\beta_{j}$.
\end{enumerate}
{ We have proposed here CV protocols for QSDC and CQD, both
of which are the schemes for secure direct quantum communication,
but the former one enables one way secure direct communication whereas
the latter one allows the users to perform a two-way secure communication
in presence of a controller. Notice from the schematic diagrams of these schemes in Figs. \ref{Fig 3} and \ref{Fig 3-1} that if the task allotted to Charlie in CQD is performed by Bob and he does not encode his message as displacement operator $D\left(\beta\right)$ the scheme is equivalent to the QSDC scheme. This reduction can be summarized as feasibility of a less complex cryptographic task if a secure communication solution is available for more complex problem. Analogously, the feasibility a of CQD scheme also provides corresponding quantum schemes for controlled DSQC, QD, DSQC as well as quantum key distribution and agreement (see \cite{thapliyal2017quantum} for more detail).}

{ Interestingly, some of the schemes for secure direct quantum communication can be used as a primitive to obtain solution of secure multiparty computation problems, which have significance in several socioeconomic tasks. For instance, consider that Alice and Bob wish to compare their assets to know the richer one among them, which is known as socialist millionaire problem \cite{saxena2019continuous}. They can perform CQD scheme where both Alice and Bob would encode the amount of assets they have and send the encoded mode finally to Charlie for measurement. Assuming that Alice (Bob) always encodes ${A}$ $\left(-{B}\right)$ for her (his) amount of assets ${A}$ $\left({B}\right)$. Depending upon whether Charlie's homodyne measurement finally gives him positive (negative) value, they can conclude Alice (Bob) the richer person as $A>B$ $\left({B}>A\right)$. Recently, 
we have used two-mode squeezed state based CQD scheme to provide
solutions for the socialist millionaire problem \cite{saxena2019continuous}. Further, these schemes ensure the feasibility of quantum
solutions for e-commerce \cite{thapliyal2019quantum} and voting \cite{thapliyal2016protocols}
as well.}
In what follows, the security of the QSDC and CQD 
schemes against various types of attacks will be critically analyzed. 

\section{Security analysis of the protocols \label{sec:Security-Analysis-of}}

  In this section, we discuss the security of the protocol
with respect to two attacks, namely the Gaussian quantum cloning machine
(GQCM) and intercept-resend attacks. 

\subsection{Gaussian quantum cloning machine (GQCM) attack \label{subsec:Gaussian-Quantum-Cloning}}

\begin{figure}
\centering{}\includegraphics[scale=0.3]{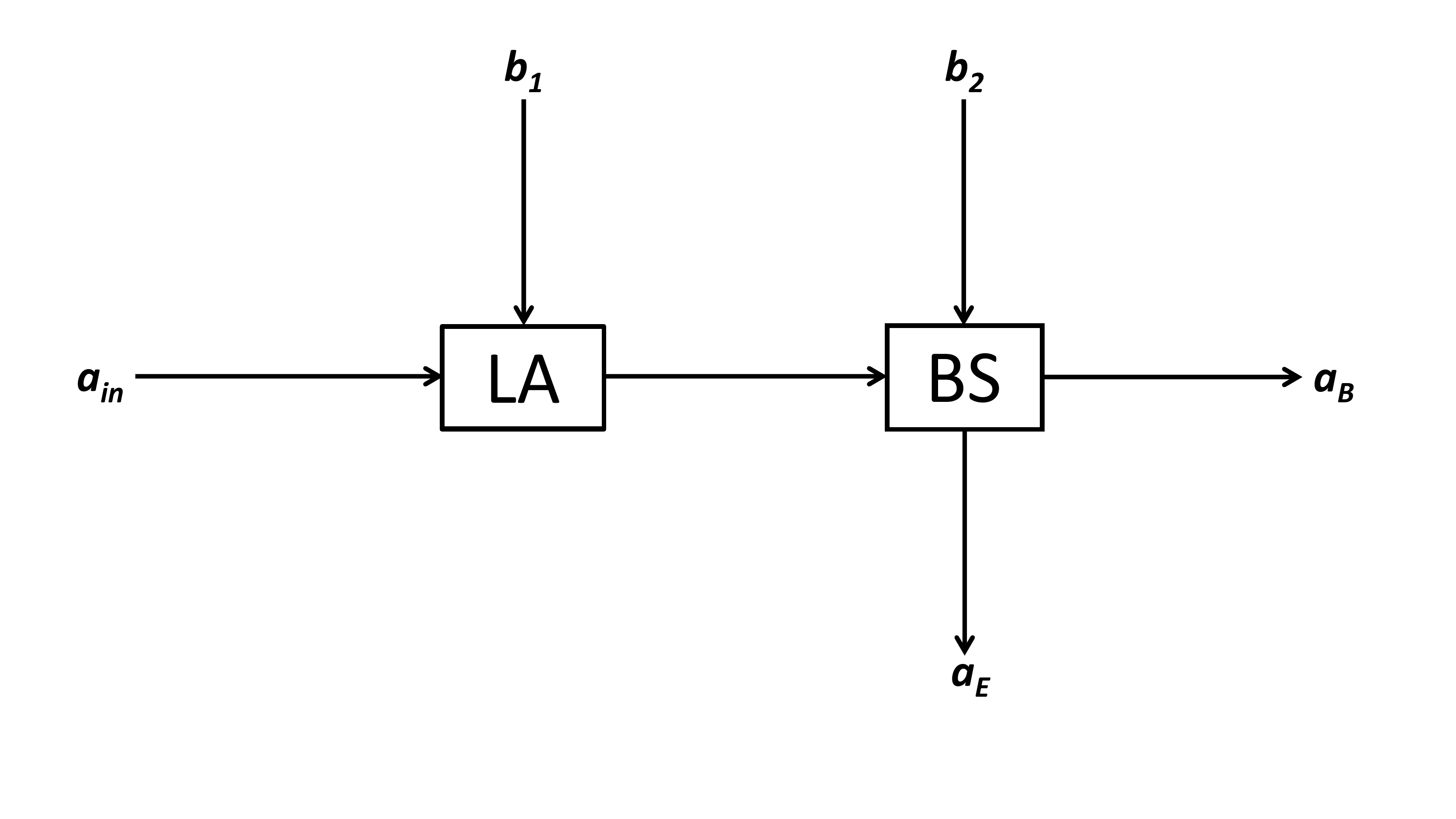}\caption{\label{fig:GC} (Color online) { Schematic diagram of the Gaussian quantum cloning machine (GQCM) attack on the CQD protocol. Here, LA and BS correspond to linear amplifier and beamsplitter, respectively.}}
\end{figure}

  A GQCM consists of a linear amplifier (LA) of gain $A$ and a
beamsplitter of transmission coefficient $T$ (shown schematically in Fig. \ref{fig:GC}). The outputs of the
GQCM are given in the Heisenberg picture as \cite{zhang2019quantum}
\begin{equation}
a_{B}=\sqrt{AT}a_{in}+\sqrt{(A-1)T}b_{1}+\sqrt{1-T}b_{2}\label{eq:GQCM1}
\end{equation}
and
\begin{equation}
a_{E}=\sqrt{T}b_{2}-\sqrt{A(1-T)}a_{in}-\sqrt{(A-1)(1-T)}b_{1},\label{eq:GQCM2}
\end{equation}
where $a_{B}$ and $a_{E}$ are Bob's and Eve's copies,
respectively. Further, $a_{in}$ is the incoming Alice's signal mode,
and \textbf{$b_{1}$} and $b_{2}$ are vacuum modes as inputs of the
linear amplifier and the beamsplitter, respectively. We discuss the
GQCM attack for the two protocols separately.

\subsubsection{QSDC protocol \label{subsec:For-the-QSDC}}

  In the QSDC protocol given in Fig. \ref{Fig 1}, the squeezing
parameters $s_{j}$ and the initial coherent amplitudes $\alpha_{j}$
are completely randomized and are kept secret by Bob throughout the
protocol. Therefore, Eve cannot infer anything from her copy of the
cloned state. Hence, the QSDC protocol is unconditionally secure against
the GQCM attack.

\subsubsection{CQD protocol \label{subsec:For-the-CQD}}

  In the first quantum channel from Charlie to Alice, Eve
has no advantage by performing this attack as Alice has not yet encoded
her message. However though, in the quantum channel from Alice to
Bob, Eve can use the GQCM on the channel and clone the incoming state,
keep her copy of the clone in a quantum memory, wait for the values
of $w_{j},\,s_{j,}\,\alpha_{j}+\beta_{j}$, and $\gamma_{j}$ to be
announced during the protocol. She, subsequently, using this information,
can measure her mode and interpret Alice's message which will reveal
Bob's message too. Similarly, an untrusted Charlie may also perform
the exact same attack.

  The input of the GQCM can be defined in Heisenberg picture
as

\begin{equation}
a_{in}=S^{\dagger}(t_{j})D^{\dagger}(d_{j})aD(d_{j})S(t_{j})=S^{\dagger}(t_{j})aS(t_{j})+d_{j},
\end{equation}
where $a=X_{a}+iP_{a}$ is the annihilation operator for the CV mode
used for message transmission. Here, $X_{a}$ and $P_{a}$ correspond
to the conjugate position and momentum quadrature operators, respectively.
The final form of displacement and squeezing operators can be defined
in terms of Alice's and Charlie's operations $D(d_{j})S(t_{j})=S(w)D(\alpha_{j})S(s_{j})D(\gamma_{j})$,
where $t_{j}=s_{j}+w_{j}=ge^{ih},$ and $d_{j}=d_{jx}+id_{jy}$ can
be obtained from $\alpha_{j}$ and $\gamma_{j}$ using Eqs. (\ref{eq:comm2})
and (\ref{eq:comm3}). 

  The use of Bogoliubov transformations yields

\begin{equation}
a_{in}=(mX_{a}+d_{jx})+i(nP_{a}+d_{jy})=X_{in}+iP_{in},\label{eq:input}
\end{equation}

  where

\[
\begin{array}{lcl}
X_{in} & = & m_{r}X_{a}-n_{i}P_{a}+d_{jx},\\
P_{in} & = & m_{i}X_{a}+n_{r}P_{a}+d_{jy}
\end{array}
\]
with 
\[
\begin{array}{lcl}
m & = & {\rm cosh}(g)-e^{ih}{\rm sinh}(g)=m_{r}+im_{i},\\
n & = & {\rm cosh}(g)+e^{ih}a^{\dagger}{\rm sinh}(g)=n_{r}+in_{i}.
\end{array}
\]

  Using Eq. (\ref{eq:input}) as the input of Eqs. (\ref{eq:GQCM1})
and (\ref{eq:GQCM2}), the position and momentum quadratures of the
Bob's and Eve's modes can be obtained as 
\begin{equation}
\begin{array}{lcl}
X_{B} & = & \sqrt{AT}(m_{r}X_{a}-n_{i}P_{a}+d_{jx})+\sqrt{(A-1)T}X_{b_{1}}+\sqrt{1-T}X_{b_{2}},\\
X_{E} & = & \sqrt{T}X_{b_{2}}-\sqrt{A(1-T)}(m_{r}X_{a}-n_{i}P_{a}+d_{jx})-\sqrt{(A-1)(1-T)}X_{b_{1}},\\
P_{B} & = & \sqrt{AT}(m_{i}X_{a}+n_{r}P_{a}+d_{jy})+\sqrt{(A-1)T}P_{b_{1}}+\sqrt{1-T}P_{b_{2}},\\
P_{E} & = & \sqrt{T}P_{b_{2}}-\sqrt{A(1-T)}(m_{i}X_{a}+n_{r}P_{a}+d_{jy})-\sqrt{(A-1)(1-T)}P_{b_{1}}.
\end{array}\label{eq:quad-out}
\end{equation}

  Since these variables are normally distributed, their corresponding
variances are 
\begin{equation}
\begin{array}{lcl}
\langle(\Delta X_{B})^{2}\rangle & = & M_{XB}+N_{XB},\\
\langle(\Delta X_{E})^{2}\rangle & = & M_{XE}+N_{XE},\\
\langle(\Delta P_{B})^{2}\rangle & = & M_{PB}+N_{PB},\\
\langle(\Delta P_{E})^{2}\rangle & = & M_{PE}+N_{PE},
\end{array}\label{eq:vari-out}
\end{equation}

  where the parameters
\[
\begin{array}{lcl}
M_{XB} & = & M_{PB}=\frac{1}{4}AT,\\
M_{XE} & = & M_{PE}=\frac{1}{4}A(1-T)
\end{array}
\]

  are the respective signal variances, and 
\[
\begin{array}{lcl}
N_{XB} & = & \frac{1}{4}((m_{r}^{2}+n_{i}^{2})AT+(A-1)T+(1-T)),\\
N_{XE} & = & \frac{1}{4}(T+(m_{r}^{2}+n_{i}^{2})A(1-T)+(A-1)(1-T)),\\
N_{PB} & = & \frac{1}{4}((m_{i}^{2}+n_{r}^{2})AT+(A-1)T+(1-T)),\\
N_{PE} & = & \frac{1}{4}(T+(m_{i}^{2}+n_{r}^{2})A(1-T)+(A-1)(1-T))
\end{array}
\]

  are the respective noise variances.

  If $\gamma_{j}$ is real (imaginary), then the position
(momentum) quadrature is used for Homodyne measurement. Therefore,
the mutual information $I_{J}(A,B)$ between Alice and Bob and $I_{J}(A,E)$
between Alice and Eve can be calculated as

\begin{equation}
I_{J}(A,B)=\frac{1}{2}\log_{2}\left(1+\frac{M_{JB}}{N_{JB}}\right),\label{eq:IAB}
\end{equation}

  and

\begin{equation}
I_{J}(A,E)=\frac{1}{2}\log_{2}\left(1+\frac{M_{JE}}{N_{JE}}\right)\label{eq:IAE}
\end{equation}
with $J\in\left\{ X,P\right\} $ corresponding to the position and
momentum quadrature measurements. The security criterion for the CQD
protocol is

\begin{equation}
\Delta I_{J}=I_{J}(A,B)-I_{J}(A,E)>0.\label{eq:sec-crit}
\end{equation}

\begin{figure}
\begin{centering}
\includegraphics[scale=0.5]{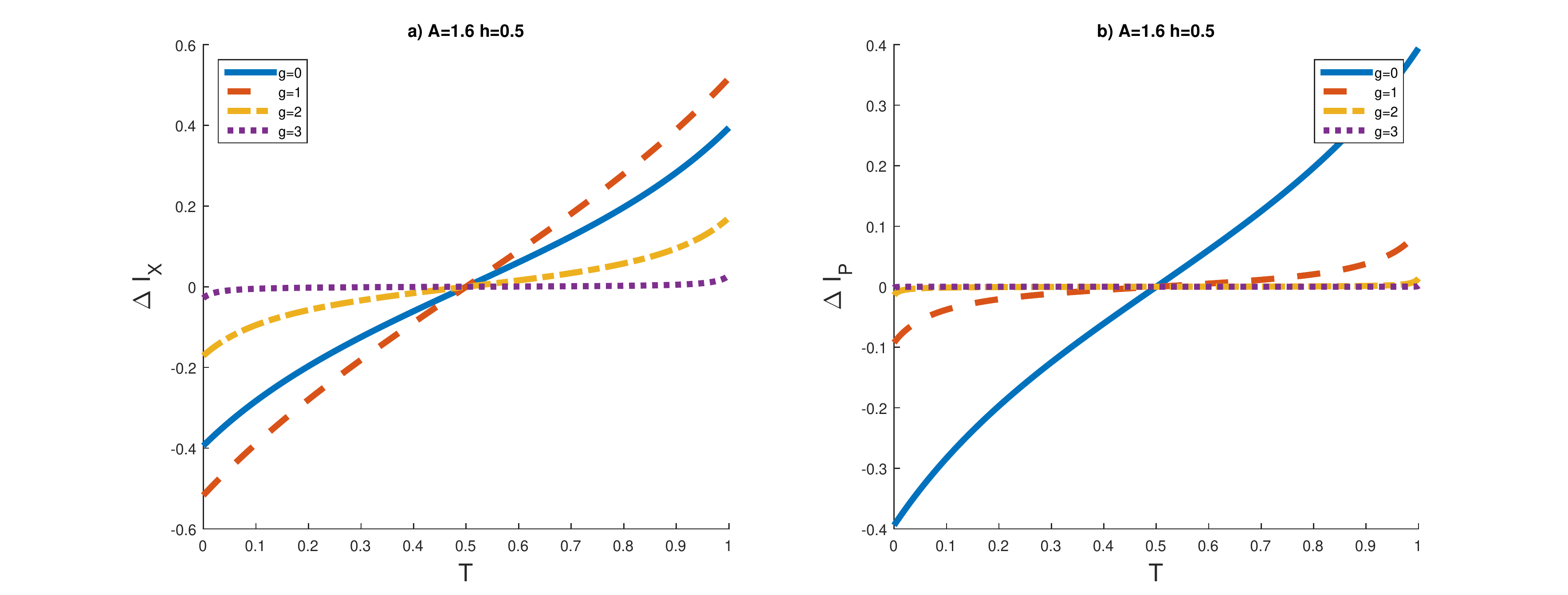}
\par\end{centering}
\centering{}\caption{\label{fig8}(Color online) Variation of (a) $\Delta I_{X}$ and (b)
$\Delta I_{P}$ with transmission coefficient $T$ for different values
of squeezing parameter $g$, while values of squeezing parameter $h$
and linear amplifier gain $A$ are mentioned. }
\end{figure}

\begin{figure}
\begin{centering}
\includegraphics[scale=0.5]{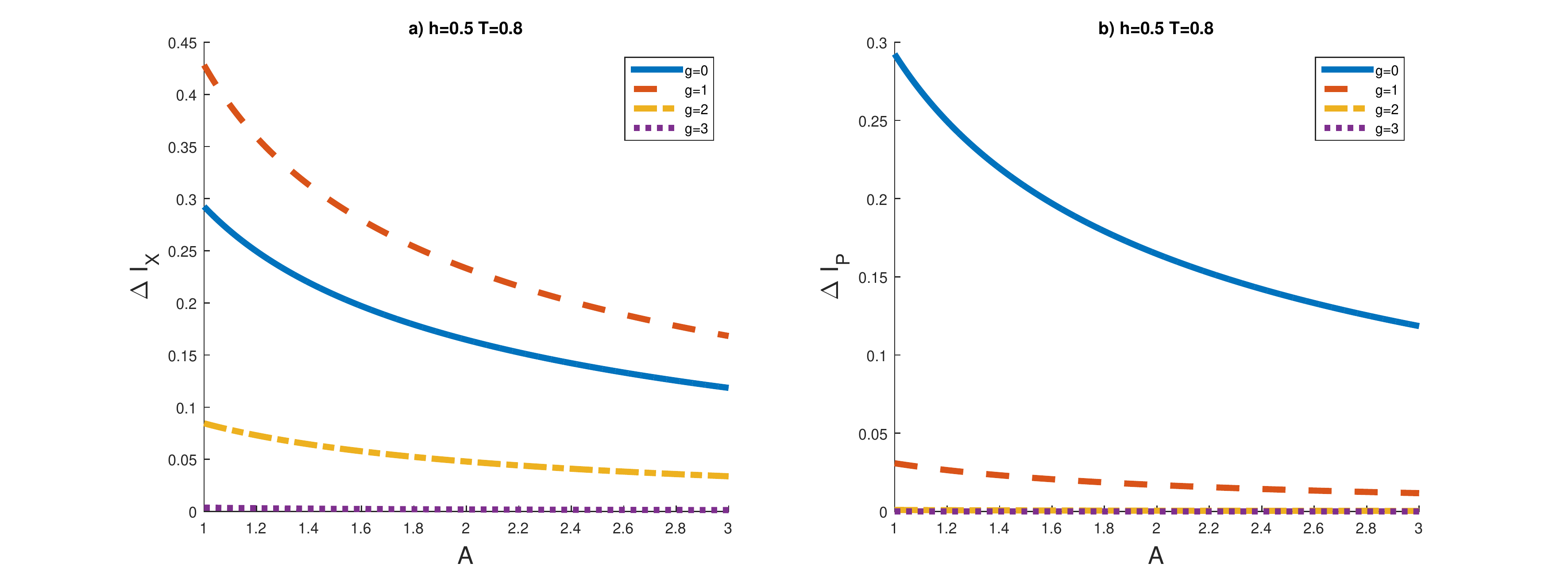}
\par\end{centering}
\centering{}\caption{\label{fig7}(Color online) Variation of (a) $\Delta I_{X}$ and (b)
$\Delta I_{P}$ with $A$ for different values of squeezing parameter
$g$, while values of $h$ and $T$ are mentioned. }
\end{figure}

\begin{figure}
\begin{centering}
\includegraphics[scale=0.5]{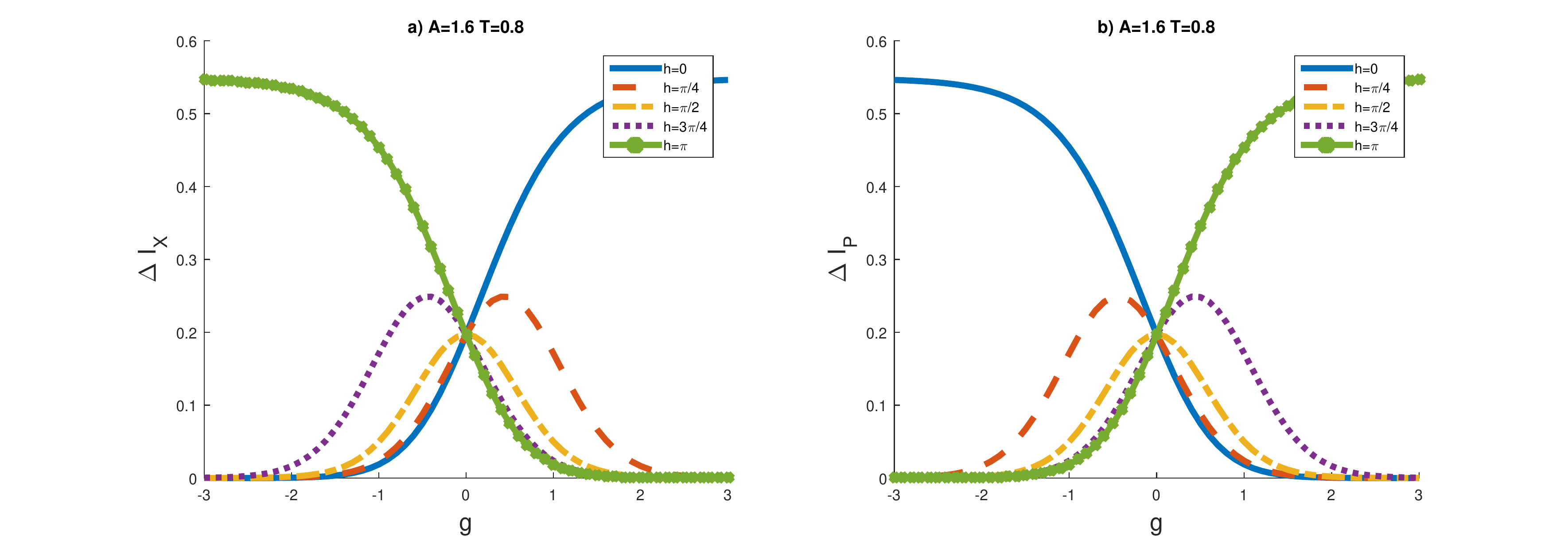}
\par\end{centering}
\caption{\label{fig5}(Color online) The security of the scheme in terms of
parameters (a) $\Delta I_{X}$ and (b) $\Delta I_{P}$ as a function
of squeezing parameters $g$ and $h$. Linear amplifier gain $A$
and transmission coefficient $T$ of the beamsplitter are mentioned. }
\end{figure}

\begin{figure}
\begin{centering}
\includegraphics[scale=0.5]{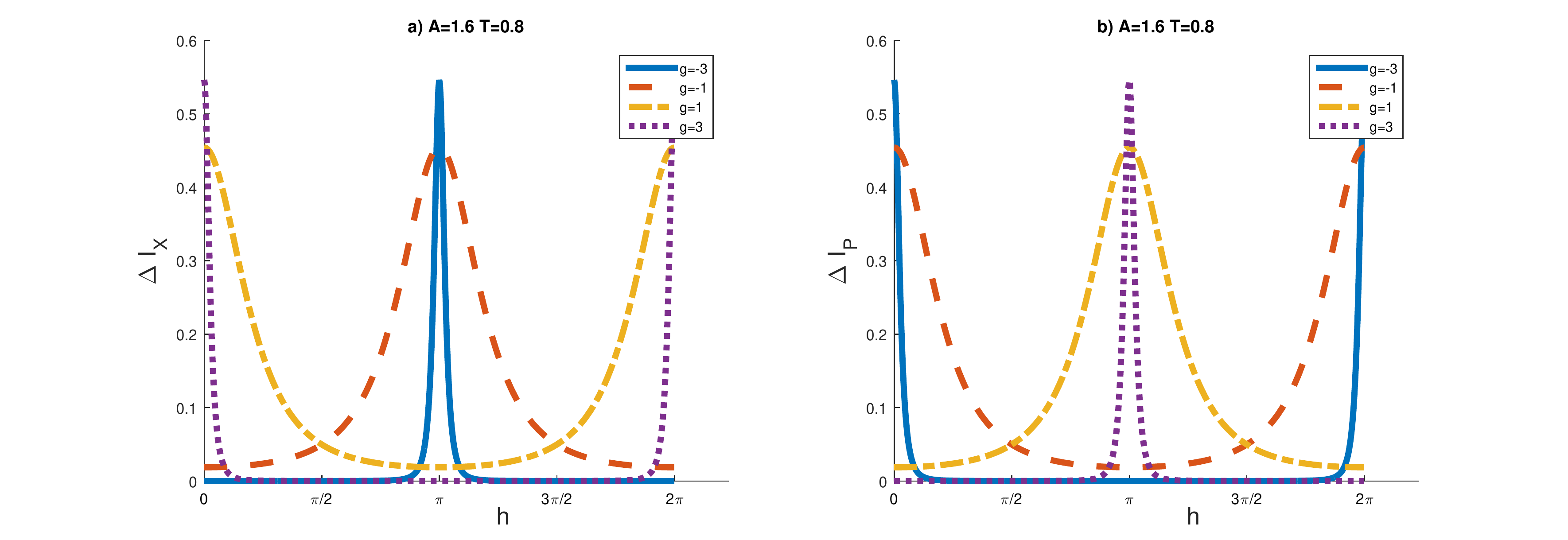}
\par\end{centering}
\caption{\label{fig6}(Color online) Variation of (a) $\Delta I_{X}$ and (b)
$\Delta I_{P}$ squeezing parameters $g$ and $h$ with the value
of $A$ and $T$ as mentioned. }
\end{figure}

We further discuss the dependence of the security of the protocol
on different parameters, such as squeezing parameters and cloning
parameters (shown in Figs. \ref{fig8}-\ref{fig6}). According to
Fig. \ref{fig8}, we can clearly see that for $\Delta I_{J}>0$ requires
the transmission coefficient $T>0.5$, and we also see that both $\Delta I_{X}$
and $\Delta I_{P}$ increases with an increase in $T$. Additionally,
for $T>0.5$, $\Delta I_{J}$ decreases with increase in the gain
parameter $A$ as shown in Fig. \ref{fig7}. In the case of $T>0.5$,
both $\Delta I_{X}$ and $\Delta I_{P}$ initially increase with increase
in the value of squeezing parameter $g$ and then decrease after reaching
a maximum value of $\Delta I_{J}$. The effect of squeezing parameter
$g$ on position and momentum quadratures is opposite as $\Delta I_{P}$
can be obtained from $\Delta I_{X}$ by considering $-g$ instead
of $g$ (cf. Figs. \ref{fig5} and \ref{fig6}). We have used $h$
in the range $0$ to $\pi$ because according to Fig. \ref{fig6},
the results are symmetric about $h=\pi$.

\subsection{Intercept-resend attack \label{subsec:Intercept-Resend-Attack}}

  In the QSDC protocol, Eve keeps the incoming state from
Bob by using delay, and she prepares and sends her own freshly prepared
state to Alice. Alice encodes her message on Eve's state and sends
it to Bob. Eve decodes the message from this state, encodes the same
message on Bob's state in the quantum memory and sends it back to
Bob. This attack cannot be performed in this protocol because Eve
will be detected when Alice checks for eavesdropping on the control
mode, as $\alpha_{j}$ values of those sent by Bob and those measured
by Eve will not match. Another type of intercept-resend attack (which
is a form of denial of service) can be performed by Eve in which she
blocks the second channel (i.e., from Alice to Bob) and prepares and
sends random quantum states to Bob, preventing Bob from receiving
the message. This attack by Eve will be detected when Bob performs
the eavesdropping checking on the decoy states in the message mode.
In another kind of attack strategy, Eve measures the intercepted state,
infers it, re-prepares it and sends it. This attack too, cannot be
performed successfully by Eve as the squeezing parameter $s_{j}$
and the initial amplitude $\alpha_{j}$ are random and unknown at
the time of measurement, and hence no information can be obtained
about the message sent. The same arguments hold for the CQD protocol.

\subsection{Participant attack by Charlie}

  In CQD protocol, after Alice records her message and sends
the encoded state to Bob. As Charlie is aware of the squeezing parameter,
so he may intercept the Alice-Bob channel and apply $S^{\dagger}$
operator to find out Alice's message. Subsequently, he re-prepares
and re-sends the state to Bob. To avoid this, Alice has to apply her
own random squeezing operation $S(w_{j})$ on the state after encoding
her message. Here, it may be noted that this attack is not applicable
to the other scheme proposed here as there is no supervisor in that
scheme. 

\section{Conclusion\label{sec:Conclusion}}

Motivated by the recent success of CV quantum key distribution schemes
using Gaussian states, we proposed here a two-party and a three-party
CV direct secure Gaussian quantum communication schemes. Specifically,
we have shown that a sender can send CV information to a distant receiver
(in the QSDC scheme) without distributing and encrypting it with a quantum
key to attain security ensured by the quantum resources. A single-mode
squeezed state is used as a quantum resource, and displacement and squeezing
operations are used to encode the message, which is decoded by performing
homodyne detection. 

We have also proposed a CV CQD which allows two parties
to perform simultaneous communication under the supervision of a controller
in a secure manner. The proposed scheme can be reduced to the corresponding
three-party scheme where a sender can send her message to the receiver
by assuming that Bob is neither encoding his message nor announcing
his final measurement outcomes in our CQD protocol. Both these schemes
are found useful as primitive to design solutions for socioeconomic
problems, and thus the feasibility of our scheme with single-mode
Gaussian states enables us to provide CV solutions
for these problems.  {We have further established the application of our CQD scheme to provide solution of socialist millionaire problem. Similarly, the proposed schemes can be used to obtain quantum
solutions for e-commerce \cite{thapliyal2019quantum} and voting \cite{thapliyal2016protocols}
as well.} Our CQD scheme can also be reduced to a modified version
of recently proposed two-party quantum dialogue \cite{zhang2019quantum}
scheme in the limiting case, which will be more robust against Gaussian
quantum cloning machine and intercept-resend attacks as the squeezing
parameter is not revealed in our scheme.

We conclude the paper with the expectation that in view of the recent
surge of experimental direct secure quantum communication and promising
future of CV quantum communication, the present schemes
be implemented in the near future. 

\section*{Acknowledgment }

 AP acknowledges the support from Interdisciplinary Cyber Physical
Systems (ICPS) programme of the Department of Science and Technology
(DST), India, Grant No.: DST/ICPS/QuST/Theme-1/2019/14. KT acknowledges
the financial support from the Operational Programme Research, Development
and Education - European Regional Development Fund project no. CZ.02.1.01/0.0/0.0/16\_019/0000754 of the Ministry of Education, Youth and Sports of the
Czech Republic.


\begin{thebibliography}{10}
\expandafter\ifx\csname urlstyle\endcsname\relax
  \providecommand{\doi}[1]{doi:\discretionary{}{}{}#1}\else
  \providecommand{\doi}{doi:\discretionary{}{}{}\begingroup
  \urlstyle{rm}\Url}\fi
\providecommand{\bibAnnoteFile}[1]{%
  \IfFileExists{#1}{\begin{quotation}\noindent\textsc{Key:} #1\\
  \textsc{Annotation:}\ \input{#1}\end{quotation}}{}}
\providecommand{\bibAnnote}[2]{%
  \begin{quotation}\noindent\textsc{Key:} #1\\
  \textsc{Annotation:}\ #2\end{quotation}}

\bibitem{dowling2003quantum}
Dowling, J.~P., Milburn, G.~J.: Quantum technology: the second quantum
  revolution. Philosophical Transactions of the Royal Society of London. Series
  A: Mathematical, Physical and Engineering Sciences \textbf{361}, 1655--1674
  (2003)
\input{dowling2003quantum}

\bibitem{bennett1984quantum}
Bennett, C.~H., Brassard, G.: Quantum cryptography: public key distribution and
  coin tossing. In: International Conference on Computer System and Signal
  Processing, IEEE, 1984 pp. 175--179 (1984)
\input{bennett1984quantum}

\bibitem{gisin2002quantum}
Gisin, N., Ribordy, G., Tittel, W., Zbinden, H.: Quantum cryptography. Rev.
  Mod. Phys. \textbf{74}, 145 (2002)
\input{gisin2002quantum}

\bibitem{pathak2013elements}
Pathak, A.: Elements of quantum computation and quantum communication. Taylor
  \& Francis, New York (2013)
\input{pathak2013elements}

\bibitem{shenoy2017quantum}
Shenoy-Hejamadi, A., Pathak, A., Radhakrishna, S.: Quantum cryptography: key
  distribution and beyond. Quanta \textbf{6}, 1--47 (2017)
\input{shenoy2017quantum}

\bibitem{bostrom2002deterministic}
Bostr{\"o}m, K., Felbinger, T.: Deterministic secure direct communication using
  entanglement. Phys. Rev. Lett. \textbf{89}, 187902 (2002)
\input{bostrom2002deterministic}

\bibitem{deng2003controlled}
Deng, F.-G., Long, G.-L.: Controlled order rearrangement encryption for quantum
  key distribution. Phys. Rev. A \textbf{68}, 042315 (2003)
\input{deng2003controlled}

\bibitem{hu2015experimental}
Hu, J., Yu, B., Jing, M., et~al.: Experimental quantum secure direct
  communication with single photons. Light: Science \& Applications \textbf{5},
  e16144 (2016)
\input{hu2015experimental}

\bibitem{long2007quantum}
Long, G.-l., Deng, F.-g., Wang, C., et~al.: Quantum secure direct communication
  and deterministic secure quantum communication. Frontiers of Physics in China
  \textbf{2}, 251--272 (2007)
\input{long2007quantum}

\bibitem{nguyen2004quantum}
Nguyen, B.~A.: Quantum dialogue. Phys. Lett. A \textbf{328}, 6--10 (2004)
\input{nguyen2004quantum}

\bibitem{thapliyal2015applications}
Thapliyal, K., Pathak, A.: Applications of quantum cryptographic switch:
  {various} tasks related to controlled quantum communication can be performed
  using {Bell} states and permutation of particles. Quantum Inf. Process.
  \textbf{14}, 2599--2616 (2015)
\input{thapliyal2015applications}

\bibitem{braunstein2005quantum}
Braunstein, S.~L., Van~Loock, P.: Quantum information with continuous
  variables. Rev. Mod. Phys. \textbf{77}, 513 (2005)
\input{braunstein2005quantum}

\bibitem{andersen2010continuous}
Andersen, U.~L., Leuchs, G., Silberhorn, C.: Continuous-variable quantum
  information processing. Laser \& Photonics Reviews \textbf{4}, 337--354
  (2010)
\input{andersen2010continuous}

\bibitem{weedbrook2012gaussian}
Weedbrook, C., Pirandola, S., Garc{\'\i}a-Patr{\'o}n, R., et~al.: Gaussian
  quantum information. Rev. Mod. Phys. \textbf{84}, 621 (2012)
\input{weedbrook2012gaussian}

\bibitem{hillery2000quantum}
Hillery, M.: Quantum cryptography with squeezed states. Phys. Rev. A
  \textbf{61}, 022309 (2000)
\input{hillery2000quantum}

\bibitem{PhysRevA.63.022309}
Gottesman, D., Preskill, J.: Secure quantum key distribution using squeezed
  states. Phys. Rev. A \textbf{63}, 022309 (2001)
\input{PhysRevA.63.022309}

\bibitem{reid2000quantum}
Reid, M.~D.: Quantum cryptography with a predetermined key, using
  continuous-variable {Einstein-Podolsky-Rosen} correlations. Phys. Rev. A
  \textbf{62}, 062308 (2000)
\input{reid2000quantum}

\bibitem{ralph1999continuous}
Ralph, T.~C.: Continuous variable quantum cryptography. Phys. Rev. A
  \textbf{61}, 010303 (1999)
\input{ralph1999continuous}

\bibitem{ralph2000security}
Ralph, T.~C.: Security of continuous-variable quantum cryptography. Phys. Rev.
  A \textbf{62}, 062306 (2000)
\input{ralph2000security}

\bibitem{grosshans2002continuous}
Grosshans, F., Grangier, P.: Continuous variable quantum cryptography using
  coherent states. Phys. Rev. Lett. \textbf{88}, 057902 (2002)
\input{grosshans2002continuous}

\bibitem{srikara2019continuous}
S, S., Thapliyal, K., Pathak, A.: Continuous variable {B92} quantum key
  distribution protocol using single photon added and subtracted coherent
  states. arXiv preprint arXiv:1906.07768  (2019)
\input{srikara2019continuous}

\bibitem{pirandola2008continuous}
Pirandola, S., Mancini, S., Lloyd, S., Braunstein, S.~L.: Continuous-variable
  quantum cryptography using two-way quantum communication. Nature Physics
  \textbf{4}, 726 (2008)
\input{pirandola2008continuous}

\bibitem{yuan2015continuous}
Yuan, L., Chunlei, J., Shunru, J., Mantao, X.: Continuous variable quantum
  secure direct communication in non-{Markovian} channel. Int. J. Theor. Phys.
  \textbf{54}, 1968--1973 (2015)
\input{yuan2015continuous}

\bibitem{marino2006deterministic}
Marino, A.~M., Stroud~Jr, C.: Deterministic secure communications using
  two-mode squeezed states. Physical Review A \textbf{74}, 022315 (2006)
\input{marino2006deterministic}

\bibitem{zhou2017new}
Zhou, N.-R., Li, J.-F., Yu, Z.-B., Gong, L.-H., Farouk, A.: New quantum
  dialogue protocol based on continuous-variable two-mode squeezed vacuum
  states. Quantum Inf. Process. \textbf{16}, 4 (2017)
\input{zhou2017new}

\bibitem{zhang2019quantum}
Zhang, M.-H., Cao, Z.-W., He, C., Qi, M., Peng, J.-Y.: Quantum dialogue
  protocol with continuous-variable single-mode squeezed states. Quantum Inf.
  Process. \textbf{18}, 83 (2019)
\input{zhang2019quantum}

\bibitem{yu2016efficient}
Yu, Z.-B., Gong, L.-H., Zhu, Q.-B., Cheng, S., Zhou, N.-R.: Efficient
  three-party quantum dialogue protocol based on the continuous variable {GHZ}
  states. Int. J. Theor. Phys. \textbf{55}, 3147--3155 (2016)
\input{yu2016efficient}

\bibitem{gong2018quantum}
Gong, L.-H., Tian, C., Li, J.-F., Zou, X.: Quantum network dialogue protocol
  based on continuous-variable {GHZ} states. Quantum Inf. Process. \textbf{17},
  331 (2018)
\input{gong2018quantum}

\bibitem{saxena2019continuous}
Saxena, A., Thapliyal, K., Pathak, A.: Continuous variable controlled quantum
  dialogue and secure multiparty quantum computation. arXiv preprint
  arXiv:1902.00458  (2019)
\input{saxena2019continuous}

\bibitem{thapliyal2016protocols}
Thapliyal, K., Sharma, R.~D., Pathak, A.: Protocols for quantum binary voting.
  Int. J. Quantum Inf. p. 1750007 (2017)
\input{thapliyal2016protocols}

\bibitem{shukla2017semi}
Shukla, C., Thapliyal, K., Pathak, A.: Semi-quantum communication: protocols
  for key agreement, controlled secure direct communication and dialogue.
  Quantum Inf. Process. \textbf{16}, 295 (2017)
\input{shukla2017semi}

\bibitem{sharma2017quantumauction}
Sharma, R.~D., Thapliyal, K., Pathak, A.: Quantum sealed-bid auction using a
  modified scheme for multiparty circular quantum key agreement. Quantum Inf.
  Process. \textbf{16}, 169 (2017)
\input{sharma2017quantumauction}

\bibitem{thapliyal2018orthogonal}
Thapliyal, K., Sharma, R.~D., Pathak, A.: Orthogonal-state-based and
  semi-quantum protocols for quantum private comparison in noisy environment.
  Int. J. Quantum Inf. \textbf{16}, 1850047 (2018)
\input{thapliyal2018orthogonal}

\bibitem{thapliyal2019quantum}
Thapliyal, K., Pathak, A.: Quantum e-commerce: a comparative study of possible
  protocols for online shopping and other tasks related to e-commerce. Quantum
  Inf. Process. \textbf{18}, 191 (2019)
\input{thapliyal2019quantum}

\bibitem{yao1982protocols}
Yao, A.~C.: Protocols for secure computations. in: Foundations of Computer
  Science, 1982. SFCS'08. 23rd Annual Symposium on pp. 160--164 IEEE (1982)
\input{yao1982protocols}

\bibitem{villar2005generation}
Villar, A. d.~S., Cruz, L., Cassemiro, K.~N., Martinelli, M., Nussenzveig, P.:
  Generation of bright two-color continuous variable entanglement. Phys. Rev.
  Lett. \textbf{95}, 243603 (2005)
\input{villar2005generation}

\bibitem{lassen2009continuous}
Lassen, M., Leuchs, G., Andersen, U.~L.: Continuous variable entanglement and
  squeezing of orbital angular momentum states. Phys. Rev. Lett. \textbf{102},
  163602 (2009)
\input{lassen2009continuous}

\bibitem{dos2009continuous}
Dos~Santos, B.~C., Dechoum, K., Khoury, A.: Continuous-variable
  hyperentanglement in a parametric oscillator with orbital angular momentum.
  Phys. Rev. Lett. \textbf{103}, 230503 (2009)
\input{dos2009continuous}

\bibitem{liu2014experimental}
Liu, K., Guo, J., Cai, C., Guo, S., Gao, J.: Experimental generation of
  continuous-variable hyperentanglement in an optical parametric oscillator.
  Phys. Rev. Lett. \textbf{113}, 170501 (2014)
\input{liu2014experimental}

\bibitem{thapliyal2014higher}
Thapliyal, K., Pathak, A., Sen, B., Pe{\v{r}}ina, J.: Higher-order
  nonclassicalities in a codirectional nonlinear optical coupler: Quantum
  entanglement, squeezing, and antibunching. Physical Review A \textbf{90},
  013808 (2014)
\input{thapliyal2014higher}

\bibitem{thapliyal2014nonclassical}
Thapliyal, K., Pathak, A., Sen, B., Perina, J.: Nonclassical properties of a
  contradirectional nonlinear optical coupler. Physics Letters A \textbf{378},
  3431--3440 (2014)
\input{thapliyal2014nonclassical}

\bibitem{thapliyal2017comparison}
Thapliyal, K., Samantray, N.~L., Banerji, J., Pathak, A.: Comparison of lower-
  and higher-order nonclassicality in photon added and subtracted squeezed
  coherent states. Phys. Lett. A \textbf{381}, 3178 -- 3187 (2017)
\input{thapliyal2017comparison}

\bibitem{agarwal2013quantum}
Agarwal, G.~S.: Quantum Optics. Cambridge University Press, Cambridge (2013)
\input{agarwal2013quantum}

\bibitem{wigner1932quantum}
Wigner, E.~P.: On the quantum correction for thermodynamic equilibrium. Phys.
  Rev. \textbf{40}, 749 (1932)
\input{wigner1932quantum}

\bibitem{strawberryfields}
Strawberry fields interactive. https://strawberryfields.ai. Xanadu, { Accessed
  on 10 December 2018}
\input{strawberryfields}

\bibitem{thapliyal2015quasiprobability}
Thapliyal, K., Banerjee, S., Pathak, A., Omkar, S., Ravishankar, V.:
  Quasiprobability distributions in open quantum systems: {Spin-qubit} systems.
  Ann. Phys. \textbf{362}, 261--286 (2015)
\input{thapliyal2015quasiprobability}

\bibitem{thapliyal2016tomograms}
Thapliyal, K., Banerjee, S., Pathak, A.: Tomograms for open quantum systems: in
  (finite) dimensional optical and spin systems. Ann. Phys. \textbf{366},
  148--167 (2016)
\input{thapliyal2016tomograms}


\bibitem{thapliyal2017quantum}
Thapliyal, K., Pathak, A., Banerjee, S.: Quantum cryptography over non-{Markovian} channels. Quantum Inf. Process.
  \textbf{16}, 115 (2017)
\input{thapliyal2015applications}

\end{thebibliography}
\end{document}